\documentclass[useAMS,usenatbib,usegraphicx]{mn2e}
\usepackage{graphicx}
\usepackage[breaklinks,colorlinks,citecolor=black,linkcolor=black,urlcolor=black]{hyperref}

\newcommand*{\mysub}[2]{\ensuremath{#1_{\mathrm{#2}}}}
\newcommand*{\unit}[1]{\ensuremath{\mathrm{\, #1}}}

\newcommand*{\Msun}{\ensuremath{\, M_{\odot}}}

\newcommand*{\erg}{\unit{erg}}

\newcommand*{\second}{\unit{s}}

\newcommand*{\LCDM}{\ensuremath{\Lambda}CDM}
\newcommand*{\Omegam}{\mysub{\Omega}{m}}

\newcommand*{\rs}{\mysub{r}{s}}
\newcommand*{\rout}{\mysub{r}{out}}
\newcommand*{\Lcool}{\mysub{L}{cool}}

\newcommand*{\E}[1]{\ensuremath{\times 10^{#1}}}
\newcommand*{\ltsim}{\ {\raise-.75ex\hbox{$\buildrel<\over\sim$}}\ }
\newcommand*{\gtsim}{\ {\raise-.75ex\hbox{$\buildrel>\over\sim$}}\ }
\newcommand*{\proptosim}{\ {\raise-.75ex\hbox{$\buildrel\propto\over\sim$}}\ }

\newcommand*{\Chandra}{{\it Chandra}}

\defcitealias{Mantz1502.06020}{I}
\newcommand*{\morphpaper}{\citetalias{Mantz1502.06020}}
\defcitealias{Mantz1402.6212}{II}
\newcommand*{\cosmopaper}{\citetalias{Mantz1402.6212}}
\defcitealias{Mantz1509.01322}{III}
\newcommand*{\profpaper}{\citetalias{Mantz1509.01322}}
\defcitealias{Applegate1509.02162}{IV}
\newcommand*{\calpaper}{\citetalias{Applegate1509.02162}}

\begin{document}

\title[Relaxed Clusters: Consistency with CDM]{Cosmology and Astrophysics from Relaxed Galaxy Clusters V: Consistency with Cold Dark Matter Structure Formation}

%\newauthor starts a new line
\author[A. B. Mantz et al.]{A. B. Mantz,$^{1,2}$\thanks{E-mail: \href{mailto:amantz@slac.stanford.edu}{\tt amantz@slac.stanford.edu}} {}
  S. W. Allen,$^{1,2,3}$
  R. G. Morris$^{1,3}$
  \\$^1$Kavli Institute for Particle Astrophysics and Cosmology, Stanford University, 452 Lomita Mall, Stanford, CA 94305, USA\\
  $^2$Department of Physics, Stanford University, 382 Via Pueblo Mall, Stanford, CA 94305, USA\\
  $^3$SLAC National Accelerator Laboratory, 2575 Sand Hill Road, Menlo Park, CA  94025, USA
}
\date{Submitted 12 April 2016. Accepted 13 July 2016.}

\pagerange{\pageref{firstpage}--\pageref{lastpage}} \pubyear{????}
\maketitle
\label{firstpage}

\begin{abstract}
This is the fifth in a series of papers studying the astrophysics and cosmology of massive, dynamically relaxed galaxy clusters. Our sample comprises 40 clusters identified as being dynamically relaxed and hot in Papers~\morphpaper{} and \cosmopaper{} of this series. Here we use constraints on cluster mass profiles from X-ray data to test some of the basic predictions of cosmological structure formation in the Cold Dark Matter (CDM) paradigm. We present constraints on the concentration--mass relation for massive clusters, finding a power-law mass dependence with  a slope of $\kappa_m=-0.16\pm0.07$, in agreement with CDM predictions. For this relaxed sample, the relation is consistent with a constant as a function of redshift (power-law slope with $1+z$ of $\kappa_\zeta=-0.17\pm0.26$), with an intrinsic scatter of $\sigma_{\ln c}=0.16\pm0.03$. We investigate the shape of cluster mass profiles over the radial range probed by the data (typically $\sim50$\,kpc--1\,Mpc), and test for departures from the simple Navarro, Frenk \& White (NFW) form, for which the logarithmic slope of the density profile tends to $-1$ at small radii. Specifically, we consider as alternatives the generalized NFW (GNFW) and Einasto parametrizations. For the GNFW model, we find an average value of (minus) the logarithmic inner slope of $\beta=1.02\pm0.08$, with an intrinsic scatter of $\sigma_\beta=0.22\pm0.07$, while in the Einasto case we constrain the average shape parameter to be $\alpha=0.29\pm0.04$ with an intrinsic scatter of $\sigma_\alpha=0.12\pm0.04$. Our results are thus consistent with the simple NFW model on average, but we clearly detect the presence of intrinsic, cluster-to-cluster scatter about the average.
\end{abstract}

\section{Introduction} \label{sec:intro}

The CDM paradigm, within which the majority of matter in the Universe is weakly interacting, has enjoyed great success in explaining astrophysical and cosmological data. The inclusion of CDM in the now standard \LCDM{} cosmological model makes clear predictions for, among other things, the mass function of gravitationally collapsed structures (e.g.\ \citealt{Jenkins0005260, Evrard0110246, Tinker0803.2706}), the ratio of baryonic mass to CDM in galaxy clusters (e.g.\ \citealt{Eke9708070, Kay0407058, Borgani0906.4370}), and the distribution of mass within the structures that form hierarchically in the Universe (e.g.\ \citealt{Bullock9908159, Navarro0311231, Navarro9611107, Gao0711.0746}). Observations testing the predicted mass function and cluster baryon fraction have largely validated CDM (e.g.\ \citealt{White1993Natur.366..429W, Bahcall9803277, Reiprich0111285, Ettori0211335, Ettori0904.2740, Allen0405340, Allen0706.0033, Allen1103.4829, Vikhlinin0812.2720, Mantz0909.3098, Mantz1407.4516, Rozo0902.3702, Sehgal1010.1025, Benson1112.5435, Hasselfield1301.0816, Planck1303.5080, Planck1502.01597}).

Observational tests of the distribution of mass within bound structures is in some ways more challenging, requiring spatially resolved measurements of the gravitational potential of objects whose mass is predominately dark. Nevertheless, an extensive body of work now exists, using primarily X-ray and gravitational lensing observations of galaxy clusters \citep{Vikhlinin0507092, Voigt0602373, Zhang0603275, Schmidt0610038, Mandelbaum0805.2552, Host0907.1097, Newman1209.1391, Okabe1302.2728, Du1510.08193, Merten1404.1376, Shan1502.00313, van-Uitert1506.03817}. In this work, we revisit the subject using \Chandra{} X-ray observations of a sample of massive, highly dynamically relaxed clusters. Although not representative of the cluster population at large, these systems are an ideal laboratory for deriving three-dimensional mass profiles based on observations of the intracluster medium (ICM) because departures from hydrostatic equilibrium and systematics due to projection are minimized. Both simulations and direct calibration using weak gravitational lensing indicate that the overall bias in \Chandra{} X-ray mass estimates for this sample is small ($\ltsim 10$ per cent; \citealt{Nagai0609247, Applegate1509.02162}). The specific features of the cluster mass distribution that we consider here are (1) the concentration parameter of cluster mass profiles, and its dependence on mass and redshift; and (2) the shape of mass profiles, in particular departures from the baseline model defined by \citet*{Navarro9611107} (hereafter NFW).

The selection of the sample of massive, relaxed clusters employed here is detailed in the first paper in this series (\citealt{Mantz1502.06020}; hereafter Paper~\morphpaper{}). Papers~\cosmopaper{}, \profpaper{} and \calpaper{} \citep{Mantz1402.6212, Mantz1509.01322, Applegate1509.02162} employ \Chandra{} data for this sample to respectively constrain cosmological parameters, through gas-mass fraction measurements; scaling relations and thermodynamics of the ICM; and the average bias of the X-ray mass determinations.

Section~\ref{sec:data} reviews our procedure for constraining cluster mass profiles from X-ray data, for which complete details are available in Papers~\cosmopaper{} and \profpaper{}, and introduces the specific mass models employed in this work. In Section~\ref{sec:results}, we review the specific predictions of CDM for cluster mass profile concentrations and shapes, present our results, and compare them to others in  the literature. Section~\ref{sec:conclusions} summarizes our conclusions. We assume a concordance flat \LCDM{} cosmology with $h=0.7$ and $\Omegam=0.3$ throughout. Unless otherwise noted, quoted parameter uncertainties correspond to the maximum-likelihood (i.e.\ shortest) interval enclosing 68.3 per cent posterior probability, and best-fitting values are the posterior modes.

\section{Data and Analysis} \label{sec:data}

This work employs the sample of 40 massive, dynamically relaxed galaxy clusters identified in Papers~\morphpaper{} and \cosmopaper{} (Table~\ref{tab:thetable}). For our purposes, dynamical relaxation was defined quantitatively in terms of X-ray image features, specifically the sharpness of the surface brightness peak and the alignment and symmetry of a series of standardized cluster isophotes; in addition all clusters in the sample must have an emission-weighted temperature of $>5$\,keV outside of their cores. The procedure for cleaning and fitting the \Chandra{} data for these clusters is described in detail in Paper~\cosmopaper{}; in Paper~\profpaper{} the data reduction was updated to use a newer version of the \Chandra{} calibration files (specifically {\sc caldb}\footnote{\url{http://cxc.harvard.edu/caldb/}} 4.6.2), and we use that version of the data reduction here. A complete list of the specific OBSIDs included in the analysis can also be found in that work. Note that, while the data for each of these 40 clusters can constrain the 2-parameter NFW mass model, the 3-parameter models discussed below can be constrained for only a subset of entire sample (see Sections~\ref{sec:gnfw} and \ref{sec:einasto}).

\begin{table*}
  \centering
  \caption[]{
    The galaxy cluster data set. Column [1] name, [2] redshift, [3--4] the radial range from which data is used to constrain mass profiles, [5--6] mass and concentration parameter constraints assuming the NFW profile, [7] central cooling luminosity, [8] inner slope constraints using the GNFW profile, and [9] Einasto profile shape parameter constraints. Quoted values are the medians of the marginalized posterior distributions for each parameter, and confidence intervals correspond to the 15.85th and 84.15th percentiles of the posterior (enclosing 68.3 per cent confidence).
  }
  \label{tab:thetable}
  \begin{tabular}{lcrrr@{ $\pm$ }lr@{ $\pm$ }lr@{ $\pm$ }lcc}
    \hline
    Name & Redshift & $R_\mathrm{min}$ & $R_\mathrm{max}$ & \multicolumn{2}{c}{$M_{200}$} & \multicolumn{2}{c}{$c$} & \multicolumn{2}{c}{$\Lcool$} & $\beta$ & $\alpha$ \\
    & & (kpc) & (kpc) & \multicolumn{2}{c}{($10^{14}\Msun$)} & \multicolumn{2}{c}{} & \multicolumn{2}{c}{$10^{45}\erg\second^{-1}$} & & \\
    \hline
    Abell~2029           &  0.078  &  58   &  928   &  11.3  &  0.4       &  6.2   &  0.2    &  0.836   &  0.004   &  $1.03_{-0.32}^{+0.25}$  &  $0.42_{-0.17}^{+0.20}$  \\
    Abell~478            &  0.088  &  71   &  1142  &  11.4  &  1.0       &  5.1   &  0.4    &  1.186   &  0.010   &  $1.24_{-0.07}^{+0.06}$  &  $0.31_{-0.08}^{+0.08}$  \\
    PKS~0745$-$191       &  0.103  &  60   &  1190  &  14.6  &  0.8       &  5.0   &  0.2    &  1.944   &  0.006   &  $0.90_{-0.02}^{+0.02}$  &  $0.35_{-0.02}^{+0.02}$  \\
    RX~J1524.2$-$3154    &  0.103  &  60   &  833   &  4.1   &  0.4       &  8.1   &  0.9    &  0.322   &  0.002   &  $0.55_{-0.34}^{+0.38}$  &  $0.56_{-0.13}^{+0.12}$  \\
    Abell~2204           &  0.152  &  63   &  1001  &  13.0  &  1.2       &  7.7   &  0.7    &  1.976   &  0.014   &  $0.51_{-0.23}^{+0.17}$  &  --                      \\
    RX~J0439.0+0520      &  0.208  &  54   &  857   &  5.1   &  0.8       &  7.9   &  1.3    &  0.463   &  0.009   &  $0.53_{-0.29}^{+0.36}$  &  $0.58_{-0.14}^{+0.11}$  \\
    Zwicky~2701          &  0.214  &  62   &  493   &  5.2   &  0.5       &  5.9   &  0.6    &  0.398   &  0.005   &  --                      &  --                      \\
    RX~J1504.1$-$0248    &  0.215  &  48   &  1540  &  17.0  &  1.7       &  5.8   &  0.4    &  6.084   &  0.049   &  --                      &  --                      \\
    Zwicky~2089          &  0.235  &  51   &  705   &  4.6   &  0.7       &  5.9   &  0.7    &  0.842   &  0.010   &  --                      &  --                      \\
    RX~J2129.6+0005      &  0.235  &  103  &  1646  &  8.7   &  1.2       &  6.0   &  1.0    &  0.950   &  0.023   &  $0.78_{-0.45}^{+0.34}$  &  $0.41_{-0.19}^{+0.21}$  \\
    RX~J1459.4$-$1811    &  0.236  &  147  &  1061  &  10.1  &  1.6       &  4.7   &  1.1    &  0.990   &  0.013   &  --                      &  --                      \\
    Abell~1835           &  0.252  &  101  &  992   &  17.3  &  1.3       &  4.5   &  0.3    &  3.331   &  0.021   &  $0.85_{-0.04}^{+0.04}$  &  $0.36_{-0.02}^{+0.03}$  \\
    Abell~3444           &  0.253  &  93   &  746   &  10.3  &  1.7       &  4.4   &  0.7    &  1.819   &  0.022   &  --                      &  --                      \\
    MS~2137.3$-$2353     &  0.313  &  50   &  722   &  5.0   &  0.5       &  8.0   &  0.8    &  2.355   &  0.029   &  --                      &  $0.30_{-0.12}^{+0.13}$  \\
    MACS~J0242.5$-$2132  &  0.314  &  54   &  1447  &  7.1   &  2.0       &  7.8   &  2.3    &  2.812   &  0.082   &  $1.43_{-0.32}^{+0.22}$  &  --                      \\
    MACS~J1427.6$-$2521  &  0.318  &  46   &  876   &  3.8   &  0.6       &  10.4  &  2.4    &  0.468   &  0.016   &  $0.83_{-0.53}^{+0.57}$  &  $0.56_{-0.17}^{+0.13}$  \\
    MACS~J2229.7$-$2755  &  0.324  &  55   &  1478  &  5.5   &  1.0       &  6.5   &  1.2    &  1.843   &  0.034   &  $1.53_{-0.25}^{+0.17}$  &  $0.06_{-0.03}^{+0.05}$  \\
    MACS~J0947.2+7623    &  0.345  &  53   &  1233  &  12.9  &  1.7       &  5.8   &  0.6    &  4.734   &  0.060   &  $1.32_{-0.17}^{+0.09}$  &  --                      \\
    MACS~J1931.8$-$2634  &  0.352  &  68   &  1249  &  11.6  &  1.6       &  5.3   &  0.6    &  3.856   &  0.033   &  --                      &  --                      \\
    MACS~J1115.8+0129    &  0.355  &  59   &  942   &  9.5   &  1.4       &  6.5   &  1.1    &  2.323   &  0.045   &  $1.11_{-0.32}^{+0.27}$  &  $0.28_{-0.09}^{+0.10}$  \\
    MACS~J1532.8+3021    &  0.363  &  50   &  1592  &  11.1  &  1.1       &  4.9   &  0.3    &  4.669   &  0.035   &  $1.20_{-0.32}^{+0.11}$  &  $0.15_{-0.03}^{+0.08}$  \\
    MACS~J0150.3$-$1005  &  0.363  &  60   &  637   &  4.0   &  0.8       &  7.8   &  1.9    &  0.892   &  0.031   &  --                      &  --                      \\
    MACS~J0011.7$-$1523  &  0.378  &  51   &  817   &  7.8   &  1.5       &  7.0   &  1.6    &  0.747   &  0.025   &  --                      &  $0.19_{-0.11}^{+0.25}$  \\
    MACS~J1720.2+3536    &  0.391  &  52   &  1001  &  8.1   &  1.3       &  6.9   &  1.4    &  1.238   &  0.043   &  --                      &  --                      \\
    MACS~J0429.6$-$0253  &  0.399  &  53   &  845   &  11.7  &  4.7       &  4.4   &  1.4    &  2.028   &  0.055   &  --                      &  --                      \\
    MACS~J0159.8$-$0849  &  0.404  &  106  &  1362  &  14.8  &  2.5       &  5.6   &  1.3    &  2.192   &  0.057   &  $0.77_{-0.46}^{+0.44}$  &  $0.47_{-0.20}^{+0.20}$  \\
    MACS~J2046.0$-$3430  &  0.423  &  44   &  1050  &  4.1   &  0.8       &  8.1   &  1.8    &  1.992   &  0.046   &  $1.21_{-0.55}^{+0.28}$  &  $0.16_{-0.11}^{+0.21}$  \\
    IRAS~09104+4109      &  0.442  &  50   &  538   &  8.7   &  1.6       &  6.1   &  0.9    &  2.866   &  0.050   &  --                      &  --                      \\
    MACS~J1359.1$-$1929  &  0.447  &  45   &  903   &  6.0   &  2.0       &  6.6   &  2.1    &  0.909   &  0.031   &  --                      &  $0.08_{-0.03}^{+0.06}$  \\
    RX~J1347.5$-$1145    &  0.451  &  51   &  1453  &  28.7  &  4.5       &  8.0   &  1.1    &  13.919  &  0.300   &  $1.11_{-0.29}^{+0.23}$  &  $0.22_{-0.08}^{+0.07}$  \\
    3C~295               &  0.460  &  52   &  551   &  6.0   &  1.8       &  5.9   &  1.5    &  0.974   &  0.026   &  $0.91_{-0.55}^{+0.33}$  &  --                      \\
    MACS~J1621.3+3810    &  0.461  &  57   &  919   &  7.3   &  1.0       &  7.2   &  1.3    &  1.183   &  0.027   &  --                      &  --                      \\
    MACS~J1427.2+4407    &  0.487  &  47   &  663   &  6.7   &  1.1       &  6.9   &  1.3    &  1.993   &  0.063   &  --                      &  --                      \\
    MACS~J1423.8+2404    &  0.539  &  44   &  799   &  7.2   &  0.9       &  7.0   &  0.8    &  3.551   &  0.053   &  $1.33_{-0.20}^{+0.14}$  &  $0.14_{-0.05}^{+0.08}$  \\
    SPT-CL~J2331$-$5051  &  0.576  &  26   &  1652  &  5.6   &  1.2       &  6.2   &  0.9    &  1.329   &  0.088   &  $0.48_{-0.26}^{+0.27}$  &  $0.57_{-0.14}^{+0.11}$  \\
    SPT-CL~J2344$-$4242  &  0.596  &  46   &  839   &  10.3  &  1.9       &  11.0  &  2.2    &  23.713  &  0.817   &  --                      &  --                      \\
    SPT-CL~J0000$-$5748  &  0.702  &  28   &  1802  &  8.1   &  3.3       &  4.3   &  1.4    &  2.368   &  0.124   &  --                      &  --                      \\
    SPT-CL~J2043$-$5035  &  0.723  &  43   &  912   &  6.1   &  1.1       &  4.6   &  0.9    &  3.115   &  0.089   &  --                      &  $0.16_{-0.05}^{+0.08}$  \\
    CL~J1415+3612        &  1.028  &  32   &  634   &  4.6   &  0.8       &  4.8   &  0.9    &  0.941   &  0.040   &  --                      &  --                      \\
    3C~186               &  1.063  &  28   &  511   &  5.5   &  1.3       &  4.8   &  1.1    &  2.934   &  0.119   &  --                      &  $0.52_{-0.19}^{+0.14}$  \\
    \hline
  \end{tabular}
\end{table*}

\begin{figure}
  \centering
  \includegraphics[scale=0.9]{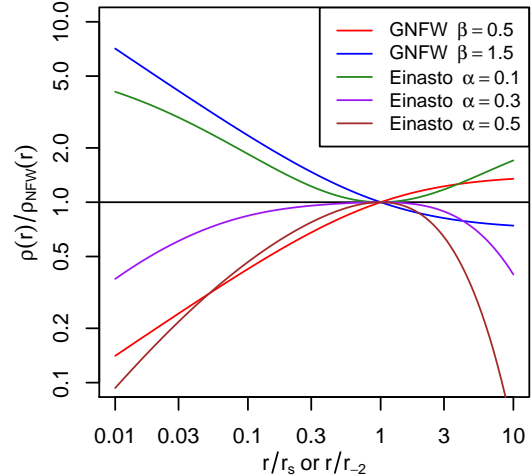}
  \caption[]{
    Comparison of the shapes of the GNFW and Einasto density profile models relative to the NFW model. Each profile is normalized at an ordinate of one. The data employed in this work are typically sensitive to radii of 0.1--1.5 in units of the NFW scale radius.
  }
  \label{fig:demo}
\end{figure}

As in our previous papers, the spectral analysis of the X-ray data here involves a deprojection, assuming spherical symmetry and hydrostatic equilibrium, in order to infer the 3-dimensional properties of the hot gas and gravitating matter in the clusters. We consider three parametric models for the total mass profiles of clusters: the standard NFW profile, whose results were previously presented in Papers~\cosmopaper{} and \profpaper{}; a generalized NFW (GNFW) profile, in which the power-law slope of the density profile at small radii is a free parameter; and the Einasto profile \citep{Einasto1969Ap......5...67E}, which provides a more accurate description of simulated halos than the other two options, particularly at small radii \citep{Navarro0810.1522}. Up to normalizing factors, the density profiles associated with these models are
\begin{eqnarray} \label{eq:profmodels}
  \mysub{\rho}{NFW}(r) &\propto& \left(\frac{r}{\rs}\right)^{-1} \left(1 + \frac{r}{\rs}\right)^{-2}, \\
  \mysub{\rho}{GNFW}(r) &\propto& \left(\frac{r}{\rs}\right)^{-\beta} \left(1 + \frac{r}{\rs}\right)^{-(3-\beta)}, \nonumber\\
  \mysub{\rho}{Einasto}(r) &\propto& \exp \left\{ -\frac{2}{\alpha}\left[\left(\frac{r}{r_{-2}}\right)^\alpha - 1\right] \right\}, \nonumber
\end{eqnarray}
where \rs{} and $r_{-2}$ are scale radii. Figure~\ref{fig:demo} shows how the shapes of the GNFW and Einasto models differ from the NFW profile. Note that, while the NFW model is a special case of the GNFW profile with $\beta=1$, no value of $\alpha$ makes the Einasto profile completely equivalent to NFW, although values of $\alpha \approx 0.2$--$0.3$ can closely approximate it over the radial range probed by our data (roughly $0.1 \ltsim r/\rs \ltsim 1.5$, in terms of the NFW scale radius).

The spectral analysis itself uses {\sc xspec}, within which we have modified the {\sc clmass} model \citep{Nulsen1008.2393} to include the GNFW and Einasto mass profiles in addition to NFW. This model treats the cluster ICM as a series of concentric, isothermal spherical shells of gas, in hydrostatic equilibrium with the gravitational potential given by one of the above parametric mass profiles. We simultaneously model absorption from Galactic hydrogen and, where necessary, the presence of foreground emission from the Galactic halo and/or the local bubble. Complete details of these aspects of the  analysis can be found in Paper~\cosmopaper{}.

For every cluster, we identify and exclude from the analysis a central region, typically 50\,kpc in radius, where visible features in the ICM indicate that the assumptions of spherical symmetry and hydrostatic equilibrium are not met. This limits the radial range over which we can draw conclusions about the shape of the mass profiles, in particular excluding the central galaxies of each cluster, where the contribution of cold gas and stars to the total mass may be significant (e.g.\ \citealt{Newman1209.1391}). The outer radius used in our analysis is set by the need to measure temperature profiles without background modelling dominating the uncertainties (see Paper~\cosmopaper{}). The specific radial range for each cluster over which the mass profiles are constrained is listed in Table~\ref{tab:thetable}.

For each cluster and each mass model considered, our spectral analysis produces a Markov chain of the parameters of interest, for which the density of samples is proportional to the joint posterior density of the parameters. Rather than summarize these results by, e.g., a mean and standard deviation, we continue to use the full set of samples when constraining properties of the ensemble of clusters (e.g. the concentration-mass relation) in Section~\ref{sec:results}. In detail, the samples for each cluster are importance weighted and then directly numerically integrated over to provide a likelihood for the hierarchical model being fitted. This is especially important for constraining the distributions of the GNFW and Einasto shape parameters, as the posterior distributions of these parameters for individual clusters are frequently poorly constrained and highly asymmetric (Table~\ref{tab:thetable}).

We additionally calculate central cooling luminosities ($\Lcool$) for each cluster in the sample. These we define as the bolometric luminosity emitted within the radius where the cooling time equals 5\,Gyr. The cooling time is calculated as
\begin{equation}
  t_\mathrm{c} = \frac{3\, n_\mathrm{tot} \,kT}{2\, n_\mathrm{e} n_\mathrm{H} \,\Lambda(T,Z)},
\end{equation}
where $T$ is the ICM temperature; $n_\mathrm{e}$, $n_\mathrm{H}$ and $n_\mathrm{tot}$ are the number densities of electrons, protons and all particles; $Z$ is the metallicity; and $\Lambda(T,Z)$ is the cooling function. These calculations are based on spectral analyses of each cluster that do not assume equilibrium or a particular underlying mass profile, using instead the {\sc projct} model in {\sc xspec}, as presented in Paper~\profpaper{}. Like total luminosity, $\Lcool$ displays an overall trend with mass; fitting a power-law plus log-normal intrinsic scatter model, we find
\begin{equation} \label{eq:Lcool-scaling}
  \frac{\Lcool}{10^{45}\erg\second^{-1}} = \left(2.9\pm0.3\right) \left(\frac{M_{200}}{10^{15}\Msun}\right)^{1.3\pm0.3},
\end{equation}
with a log-normal scatter of $0.70\pm0.09$.

\section{Results and Discussion} \label{sec:results}

This section presents our results and discusses them in the context of CDM predictions and other measurements in the literature. In Section~\ref{sec:cm}, we assume the NFW mass model and investigate the mass and redshift dependence of cluster concentrations. Sections~\ref{sec:gnfw} and \ref{sec:einasto} respectively consider the GNFW and Einasto generalizations to the NFW profile.

\subsection{NFW Concentration--Mass Relation} \label{sec:cm}

The concentration of mass within an NFW halo can be described by the parameter
\begin{equation}
  c = \frac{r_{200}}{\rs},
\end{equation}
where $r_{200}$ is the radius within which the mean enclosed density is 200 times the critical density of the Universe at a cluster's redshift. Numerous dark-matter-only simulations have shown that the concentration of halos is expected to vary with both redshift and mass on average, and to have an intrinsic scatter of $\sim20$ per cent driven primarily by differences in accretion history \citep{Navarro9611107, Bullock9908159, Zhao0309375, Zhao0811.0828, Tasitsiomi0311062, Neto0706.2919, Gao0711.0746, Duffy0804.2486, Klypin1002.3660, Klypin1411.4001, Ludlow1206.1049, Ludlow1312.0945, Prada1104.5130, Dutton1402.7073}. Early work suggested a scaling relation similar to $c \propto (1+z)^{-1} \, M^{-0.13}$ \citep{Bullock9908159}. More recent work revealed that the dependence on both mass and redshift is reduced for the most massive (i.e., cluster-scale) halos (e.g.\ \citealt{Gao0711.0746, Klypin1002.3660, Prada1104.5130}), largely because these halos are preferentially recently formed (e.g.\ \citealt{Ludlow1312.0945}). \citet{Ludlow1206.1049} have shown that selecting the most dynamically relaxed halos from simulations recovers a power-law with mass out to the largest masses. Our current understanding, at least on the basis of N-body simulations, is that concentrations for individual clusters grow steadily as they accrete through minor mergers, but that sufficiently disruptive mergers can ``reset'' the concentration parameter to lower values ($c\sim3$), with the overall trends largely reflecting differences in the typical accretion histories of halos as a function of mass and redshift \citep{Ludlow1206.1049, Klypin1411.4001}.

By construction, our sample contains only the most dynamically relaxed, massive clusters at any redshift. We therefore expect their accretion histories at the time of observation to be similar, and in particular that the time since the last major merger is significant in dynamical terms. As a consequence, we do not necessarily expect the redshift dependence found in simulations for all halos to apply to our sample, since it must partially reflect evolution in the rate of major mergers that will not be present in our sample. It therefore makes sense to consider models where both the dependence on mass and redshift are free parameters,
\begin{equation} \label{eq:cm}
  c = c_0 \, \left(\frac{1+z}{1.35}\right)^{\kappa_\zeta} \left( \frac{M_{200}}{10^{15}\Msun} \right)^{\kappa_m}.
\end{equation}
Our model also includes as a free parameter a log-normal intrinsic scatter in concentration, $\sigma_{\ln c}$, which is assumed to have the same value at all masses and redshifts.

\begin{figure*}
  \centering
  \hspace{5mm}
  \includegraphics[scale=0.86]{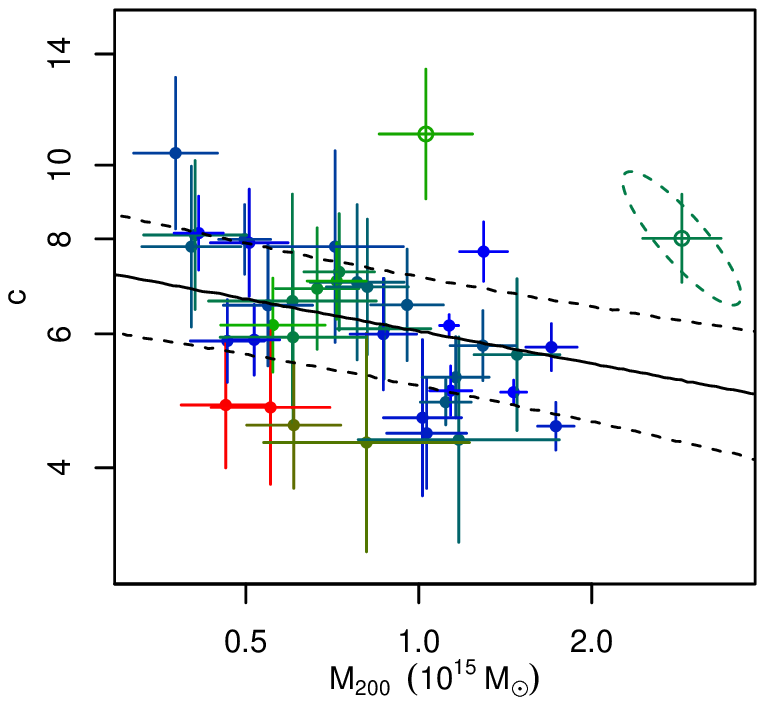}
  \hspace{2mm}
  \includegraphics[scale=0.86]{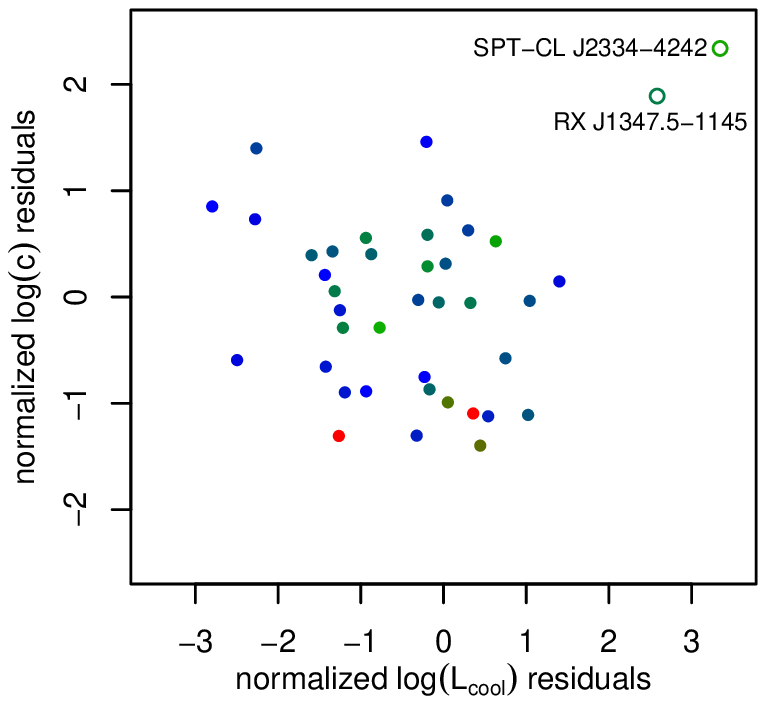}
  \includegraphics[scale=0.86]{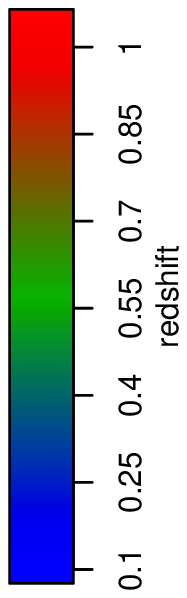}
  \caption[]{
    Left: Mass--concentration relation from the NFW analysis of the 40-cluster sample. The  fitting procedure accounts for the covariance between mass and concentration, shown explicitly for RX\,J1347.5$-$1145 (the rightmost point), although for clarity uncertainties are shown as crosses otherwise. The solid line and dashed lines respectively show the best-fitting power law in mass and the corresponding 68 per cent predictive interval, including intrinsic scatter.
    Right: Residuals in concentration are plotted against residuals in central cooling luminosity. Residuals are normalized by the combined measurement and intrinsic scatter, and are with respect to the best-fitting scaling law with $M_{200}$ in both cases.
    The clusters with the highest values of $\Lcool$, RX\,J1347.5$-$1145 and SPT-CL\,J2344$-$4242, are shown as open circles in both panels.
  }
  \label{fig:cM}
\end{figure*}

The concentrations and masses of our clusters, derived from fitting the NFW profile, are shown in the left panel of Figure~\ref{fig:cM}. Note that the measurement correlation between $M_{200}$ and $c$ is typically strong, as illustrated by a joint constraint ellipse for the case of RX\,J1347.5$-$1145, and that this correlation is accounted for by our fitting method (Section~\ref{sec:data}). Fitting the model of Equation~\ref{eq:cm}, we find $\kappa_m = -0.16 \pm 0.07$ and $\kappa_\zeta = -0.17 \pm 0.26$. A dependence on mass, consistent with the expectation from simulations, is detected at $\sim2\sigma$ significance, while the redshift dependence is consistent with zero. We find an intrinsic scatter of $\sigma_{\ln c}=0.16\pm 0.03$, somewhat smaller that the $\sim0.2$ scatter found for all halos in simulations. The results are consistent with the notion that our sample consists of highly relaxed clusters, for which we expect a power-law with mass to hold even at the highest masses, with a smaller intrinsic scatter than the full population. This is reinforced by the fact that our concentrations are on average greater than those in simulations at the same masses. Selecting relatively relaxed halos from simulations based on the absence of substructure and the ratio of kinetic to potential energy, \citet{Ludlow1206.1049} find median concentrations of 4--5 in the mass range considered here, consistent with our observed clusters being typically more relaxed than the halos they selected.\footnote{\citet{Ludlow1206.1049} classify 31 per cent of halos at $z=0$ as relaxed, 3--4 times as many as we would expect to select from a mass-limited sample based on the criteria used to construct our cluster data set (Paper~\morphpaper{}).} These authors also find a clear correlation between halo concentration and the time since ``formation'', operationally defined as the time when the halo reaches half of its final mass (see also \citealt{Wechsler0108151}). Taking their $z=0$ and $z=1$ results as a guide, our measured concentrations in the range 5--8 place the clusters' formation times approximately between redshifts 1--2. Our selection of the most relaxed systems is also plausibly responsible for a reduced redshift dependence ($\kappa_\zeta$ consistent with zero) compared with all clusters in simulations. 

The right panel of Figure~\ref{fig:cM} shows the residuals from the best-fitting concentration--mass relation as a function of central cooling luminosity (as a proxy for the importance of non-gravitational physics), with the latter also expressed as a residual from the best-fitting $\Lcool$--mass relation (Equation~\ref{eq:Lcool-scaling}). While the two sets of residuals are not strongly correlated, it may be worth noting the two clusters with large positive residuals in both concentration and cooling luminosity, RX\,J1347.5$-$1145 and SPT-CL\,J2344$-$4242. If the relatively efficient cooling at work in the centers of these clusters has an effect on the concentration of the dark matter as well as the gas, boosting the value of $c$, this would not be represented in the dark-matter-only simulations used to predict the c--M relation. (The closest analogs of these systems in simulations may, of course, still have greater than average concentration.)

Concentration--mass estimates based on X-ray data, mostly for low-redshift clusters, show broad agreement with the mass dependence predicted from N-body simulations \citep{Vikhlinin0507092, Voigt0602373, Zhang0603275, Schmidt0610038, Amodeo1604.02163}. More recently, measurements based on weak gravitational lensing have become possible. Weak lensing shear alone can constrain concentrations only for stacks or ensembles of clusters, but individual cluster constraints are possible in combination with strong lensing or magnification. Our own analysis of weak lensing data for a subset of the clusters used in this work finds an ensemble average concentration of $3.0^{+4.4}_{-1.8}$ (Paper~\calpaper{}), consistent with our X-ray results here within the statistical uncertainties. Results using strong and weak lensing by X-ray selected clusters \citep{Okabe1302.2728} and from stacked analyses of optically selected clusters \citep{Mandelbaum0805.2552,  Du1510.08193, Shan1502.00313, van-Uitert1506.03817} are in broadly good agreement with our results. However, an arguably better comparison is with the lensing analyses of the CLASH sample of clusters ($0.19<z<0.89$), which mostly consists of relaxed or nearly relaxed clusters (by our definition). Using CLASH strong+weak lensing, \citet{Merten1404.1376} found $\kappa_m=-0.32\pm0.18$  and $\kappa_\zeta=-0.14\pm0.52$, consistent with our results; incorporating magnification, \citet{Umetsu1507.04385} found $\kappa_m=-0.44\pm0.19$ while fixing $\kappa_\zeta=-0.668$.

We conclude that our concentration--mass estimates for relaxed clusters are in good agreement with simulations and previous measurements, given the expectation that we have selected dynamically similar, and, in particular, the most dynamically relaxed, clusters at all redshifts. Direct tests of that selection on simulated cluster images would not be straightforward, given the inability of current hydrodynamic simulations to reproduce the cores of relaxed clusters. However, the sensitivity of our results to a few strongly cooling clusters motivates doing so in future work, as well as expanding the relaxed cluster sample if possible.

\subsection{Generalized NFW Mass Profiles} \label{sec:gnfw}

\begin{figure*}
  \centering
  \includegraphics[scale=0.9]{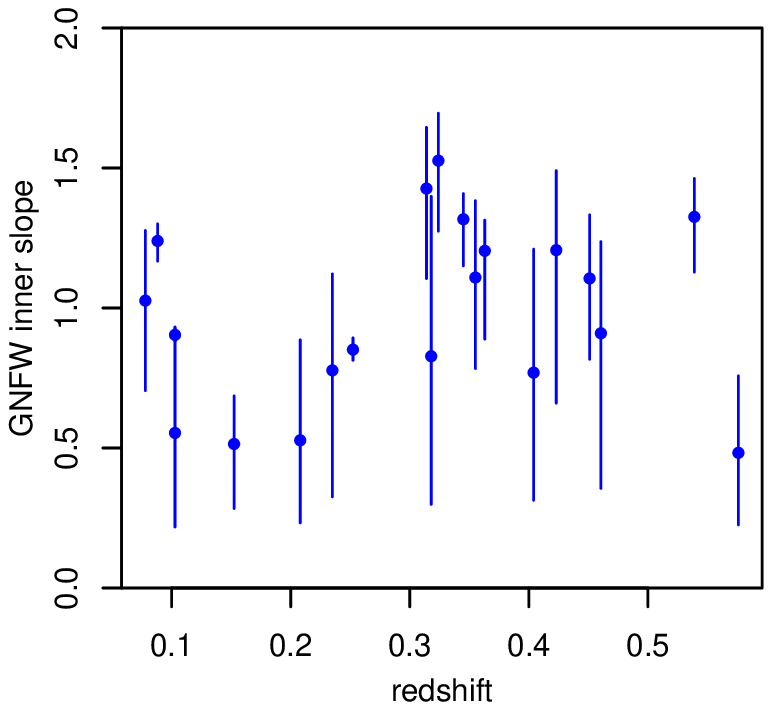}
  \hspace{1cm}
  \includegraphics[scale=0.9]{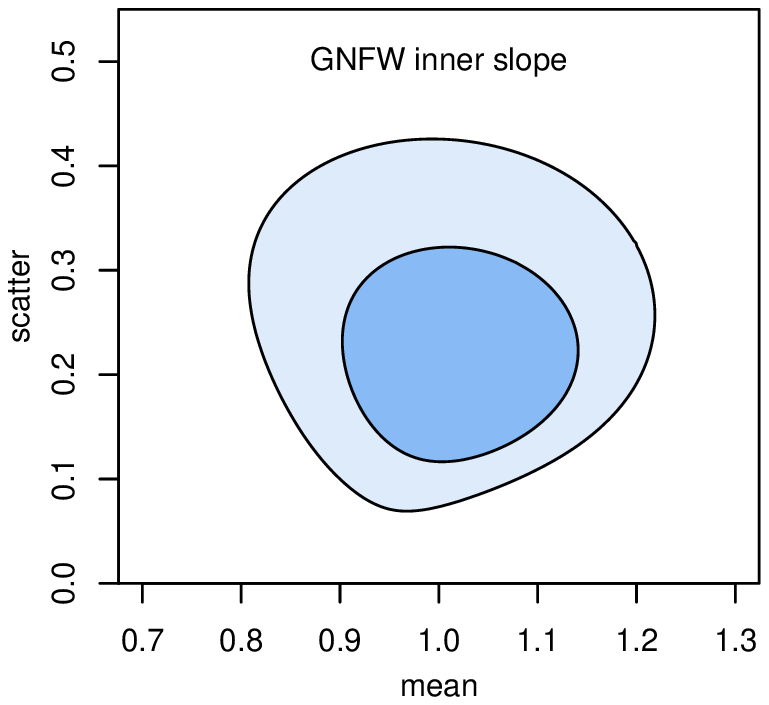}
  \caption[]{
    Left: The GNFW inner slope, $\beta$, as a function of redshift, for the subset of our sample where this model can be constrained (see Equation~\ref{eq:profmodels}). When fitting individual clusters, $\beta$ is allowed to vary between 0.0 and 2.0.
   Right: Joint 68.3 and 95.4 per cent confidence constraints on the average value and Gaussian intrinsic scatter of the inner slope parameter. The NFW model corresponds to a value of $\beta=1$.
  }
  \label{fig:gnfw}
\end{figure*}

We next turn to constraints on the inner slope of the cluster mass profiles, as parametrized by the GNFW model. While a profile of the Einasto form is now known to provide a better description of simulated clusters over a large range in radii, the GNFW model still represents a simple modification of the baseline NFW model, and one that is arguably easier to interpret in light of the limited radial range of our data. N-body simulations generally place values of the inner slope of the GNFW profile, $\beta$ (Equation~\ref{eq:profmodels}), in the range 1.0--1.5, with $\beta=1$ corresponding to the NFW model \citep{Diemand0402267, Diemand0504215, Navarro0311231, Tasitsiomi0311062}.

Unfortunately, due to its greater complexity compared to the NFW profile, the GNFW model cannot be fully constrained for all of the clusters in our sample using the current data. In practice, the scale radius parameter becomes unconstrained, taking on very large values, while the inner slope becomes relatively steep, taking values roughly appropriate for the average of the profile in the observed region (i.e., $\sim-2$). A straightforward way to eliminate these cases is to adopt the outer radius where the data can constrain the mass profile, $\rout$,  as an upper limit for the scale radius (when fitting the NFW profile, all scale radii are comfortably within the measured range). Those clusters for which the 95.4 per cent confidence interval for $\rs$ includes values $>\rout$ are removed from this portion of the analysis. We have checked that making this cut based on the 68.3 per cent confidence interval for $\rs$ does not change our results apart from tightening the constraints slightly (because fewer clusters are excluded). Thus, while this procedure at some level introduces the prior that the mass profiles not be too far from NFW, our results are not very sensitive to it. The left panel of Figure~\ref{fig:gnfw} shows these individual constraints (see also Table~\ref{tab:thetable}).

Using the 20 clusters that pass this check, we fit a model for the average value of $\beta$ and its cluster-to-cluster scatter, which we assume to be described by a Gaussian distribution with mean $\bar{\beta}$ and width $\sigma_\beta$. The joint constraints on these parameters are shown in the right panel of Figure~\ref{fig:gnfw}, and are consistent with the NFW value of $\beta=1$. The marginalized constraints are $\bar{\beta} = 1.02 \pm 0.08$ and $\sigma_\beta = 0.22 \pm 0.07$. We emphasize that the data used to fit this model do not extend all the way into the cluster centers, although they do include radii down to $\sim 0.1\,\rs$ (see Section~\ref{sec:data}).

Our analysis indicates that, on average, the NFW model provides a good description of the most relaxed clusters down to small radii. However, it also shows significant intrinsic scatter in $\beta$, at the $\sim20$ per cent level, even among the most relaxed clusters. Earlier work by \citet{Schmidt0610038} and \citet{Host0907.1097} using X-ray observations reached similar conclusions on the mean value of $\beta$, respectively finding $\bar{\beta}=0.88^{+0.26}_{-0.31}$ and $0.98<\bar{\beta}<1.19$ (both 95 per cent confidence intervals). Neither of these authors fit for intrinsic scatter explicitly, but their results also pointed to its presence, to the extent that the scatter in individual-cluster constraints on $\beta$ was not consistent with measurement uncertainties alone. More recently, \citet{Newman1209.1391} also found inner mass profile slopes of massive clusters to be consistent with CDM-only predictions, using a combination of strong and weak gravitational lensing constraints, at radii where stars do not dominate the mass budget.

\subsection{Einasto Mass Profiles} \label{sec:einasto}

\begin{figure*}
  \centering
  \includegraphics[scale=0.9]{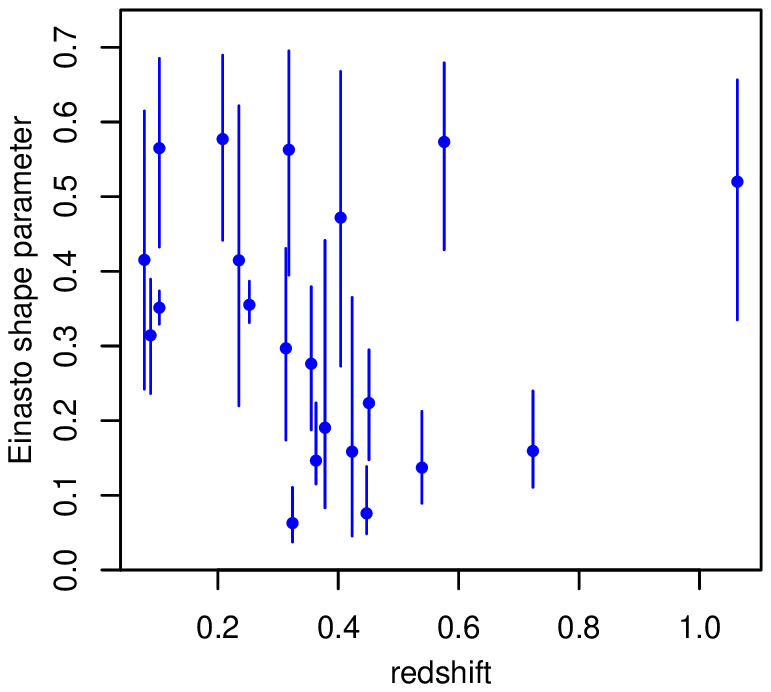}
  \hspace{1cm}
  \includegraphics[scale=0.9]{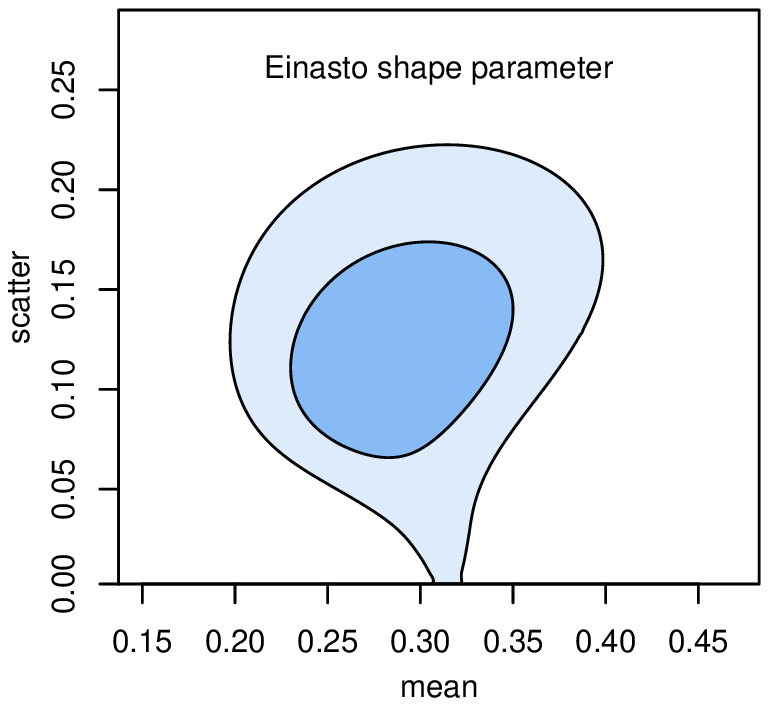}
  \caption[]{
    Left: The Einasto shape parameter, $\alpha$, as a function of redshift, for the subset of our sample where this model can be constrained (see Equation~\ref{eq:profmodels}). When fitting individual clusters, $\alpha$ is allowed to vary between 0.01 and 0.75.
    Right: Joint 68.3 and 95.4 per cent confidence constraints on the average value and Gaussian intrinsic scatter of the shape parameter. While no value of $\alpha$ perfectly corresponds to the NFW model, the values consistent with our data approximately reproduce it over the radial range probed (see Figure~\ref{fig:demo}).
  }
  \label{fig:ein}
\end{figure*}
 
In detail, the cluster mass profile shapes in simulations have been found to be better described by a continuously changing power-law, such as the Einasto form (Equation~\ref{eq:profmodels}), than by the NFW or GNFW models (e.g.\ \citealt{Navarro0810.1522}). Values of the shape parameter, $\alpha$, seen in simulations are mass and redshift dependent; for massive clusters at $z\ltsim1$, average values of $\sim0.2$--0.3 are typical \citep{Gao0711.0746, Dutton1402.7073, Klypin1411.4001}. While we do not expect our data to be capable of distinguishing the GNFW and Einasto models, due to the limited radial range covered, it is nonetheless interesting to see how values of $\alpha$ measured from the X-ray data compare to these expectations.

Similarly to the GNFW case, degeneracies between model parameters prevent us from constraining the Einasto model for every cluster in our sample. We follow the same procedure as in Section~\ref{sec:gnfw} to remove from the analysis clusters where the scale radius of the Einasto model is not constrained. Unlike the GNFW case, the constraints on $\alpha$ for several of the remaining clusters are consistent with either the lower or upper bound of the uniform prior we adopt in the analysis, $0.01\leq\alpha\leq0.75$ (left panel of Figure~\ref{fig:ein}). While this makes our results dependent on the width of this prior, in practice the effect is small; for example, our results below are nearly identical to those we would obtain with a tighter prior of $0.01\leq\alpha\leq0.5$.

As in the previous section, we fit a Gaussian model for the intrinsic distribution of Einasto shape parameters, parametrized by $\bar{\alpha}$ and $\sigma_\alpha$. The results are shown in the right panel of Figure~\ref{fig:ein}; the marginalized constraints are $\bar{\alpha} = 0.29 \pm 0.04$ and $\sigma_\alpha = 0.12 \pm 0.04$. Although the intrinsic scatter is relatively large, $\sim 40$ per cent, the mean value of the shape parameter is within the range expected for the most massive galaxy clusters at low redshifts \citep{Klypin1411.4001}. Note that, for this value of $\alpha$, the density profile is relatively similar to NFW over the range probed (Figure~\ref{fig:demo}); hence, this result is consistent with the value $\beta \approx 1$ arrived at when fitting the GNFW model.

From their analysis of X-ray data for 11 clusters, \citet{Host0907.1097} found $0.14<\bar{\alpha}<0.26$ at 95 per cent confidence, with a best-fitting value of $\sim0.2$. This is somewhat smaller than our result, although we note that their cluster sample is also lower in mass on average ($M_{200}\sim2\E{14}\Msun$), such that these results are not necessarily in conflict.

\section{Conclusion}
\label{sec:conclusions}

We present constraints on mass profile models for massive, relaxed galaxy clusters based on an analysis of X-ray data from \Chandra{}. This analysis assumes hydrostatic equilibrium between the ICM and the gravitational potential, and includes data only at cluster radii where we are confident that both departures from equilibrium and systematics due to background modeling are minimal.

Assuming the NFW mass profile model, the measured concentration--mass relation has a power-law slope with mass of $\kappa_m=-0.16\pm0.07$, consistent with CDM simulations of cosmological structure formation. The measured relation is consistent with being constant as a function of redshift, a feature that is not seen in simulations; however, this is plausibly the result of our selection of the most dynamically relaxed clusters at a given redshift. Simulations including gas physics, on which the same selection procedure is replicated, would be required to perform a completely fair comparison. We detect an intrinsic scatter of $\sigma_{\ln c}=0.16\pm0.03$ in the concentration--mass relation. Two  clear high-concentration outliers from the mean relation also have the largest central cooling luminosities in the sample, suggesting a role for baryonic physics in the scatter. However, the remaining clusters do not support a simple trend between concentration and cooling luminosity at fixed mass.

When fitting a GNFW mass profile, where the slope at small radii is a free parameter, we find an average value that is consistent with the NFW model, $\bar{\beta}=1.02\pm0.08$. However, there is significant cluster-to-cluster scatter ($\sigma_\beta=0.22\pm0.07$), even within the relaxed sample of clusters analyzed here. For the Einasto profile, we similarly measure a mean value, $\bar{\alpha}=0.29\pm0.04$, which is consistent with CDM predictions for clusters of the mass studied here. In this case, we measure a larger fractional intrinsic scatter, $\sigma_\alpha=0.12\pm0.04$.

Overall, our results confirm that the mass distribution within the most massive halos is in agreement with CDM predictions. There are clear opportunities to improve this analysis in the future, both by finding additional highly relaxed clusters and obtaining deeper X-ray data (especially to better constrain 3-parameter models like the GNFW and Einasto models). Although it is beyond the scope of this work, X-ray data for dynamically relaxed clusters at intermediate radii can potentially be combined with weak lensing (at larger radii, $\gtsim r_{500}$), and strong lensing or stellar velocity dispersions (at small radii, $\ltsim 0.5r_{2500}$), in the vein of e.g.\ \citet{Newman1209.1391}. While these additional data sets have their own challenges in the form of projection effects and velocity anisotropy, a careful combination for an appropriately chosen cluster sample potentially provides the most complete possible view of the mass distribution within clusters.

\section*{Acknowledgements}
We acknowledge support from the U.S. Department of Energy under contract number DE-AC02-76SF00515; from the National Aeronautics and Space Administration (NASA) through \Chandra{} Award Numbers GO8-9118X and TM1-12010X, issued by the \Chandra{} X-ray Observatory Center, which is operated by the Smithsonian Astrophysical Observatory for and on behalf of NASA under contract NAS8-03060; and from NASA under Grant Number NNX15AE12G.

\def \aap {A\&A} 
\def \aapr {A\&AR} 
\def \aaps {A\&AS} 
\def \statisci {Statis. Sci.} 
\def \physrep {Phys. Rep.} 
\def \pre {Phys.\ Rev.\ E} 
\def \sjos {Scand. J. Statis.} 
\def \jrssb {J. Roy. Statist. Soc. B} 
\def \pan {Phys. Atom. Nucl.} 
\def \epja {Eur. Phys. J. A} 
\def \epjc {Eur. Phys. J. C} 
\def \jcap {J. Cosmology Astropart. Phys.} 
\def \ijmpd {Int.\ J.\ Mod.\ Phys.\ D} 
\def \nar {New Astron. Rev.} 
\def \araa {ARA\&A}
\def \aj {AJ}
\def \aar {A\&AR}
\def \apj {ApJ}
\def \apjl {ApJL}
\def \apjs {ApJS}
\def \asl {Adv. Sci. Lett.} 
\def \mnras {MNRAS}
\def \nat {Nat}
\def \pasj {PASJ}
\def \pasp {PASP}
\def \science {Sci}
\def \compcom {Comput.\ Phys.\ Commun.}
\def \gca {Geochim.\ Cosmochim.\ Acta}
\def \npa {Nucl.\ Phys.\ A}
\def \plb {Phys.\ Lett.\ B}
\def \prc {Phys.\ Rev.\ C}
\def \prd {Phys.\ Rev.\ D}
\def \prl {Phys.\ Rev.\ Lett.}
\def \ssr {Space Sci.\ Rev.}

\label{lastpage}

% \bsp
\end{document}